# Prediction of USA November 2020 Election Results Using Multifactor Twitter Data Analysis Method

İbrahim SABUNCU[1], Mehmet Ali BALCI[2], Ömer AKGÜLLER[3]

## Abstract

Social media data, especially Twitter, is used to forecast election results for more than a decade. In such studies, estimates were made using one of the factors such as the number of positive, negative, and neutral tweets posted about parties, the effect size of these tweets (the number of re-tweets), or the number of people who posted these tweets. However, no study was found that used all of these factors together. The goal of this study is to develop a new approach that takes into account all of the described factors and contributes to the literature in this context. A new multifactor model for the election result prediction based on Twitter data has been developed for this purpose. This new model uses the Page Ranks centrality of users in Twitter and sentiment scores. Moreover, the forecasting is performed on the medians of multiscore distributions by using the autoregressive fractionally integrated moving average model (FARIMA). The model was tested by attempting to predict the results of the USA 2020 elections in November, which had not yet taken place when the first version of this article was written. Also, a comparison has been made with three alternative estimation approaches in the literature and polls. Analyzes were made for more than 10 million tweets collected between September 1, 2020, to November 2, 2020. As a result of the analysis, it has been observed that the newly developed multifactor method gives better results not only than other methods in the literature but also polls. The USA November 2020 presidential election results had estimated by only a 1,52% mean absolute error rate.

## 1. Introduction

Analysis of the relationship between social media data and electoral outcomes has started with the broad use of social media. Numerous studies have shown that social media has an impact on election outcomes in various countries [1] and can be used effectively to forecast election results [2–22].

Twitter is the most popular social media platform used for the prediction of election result studies using social media data. [23]. Twitter is a social media application in many countries where millions of people share their views on different topics, politicians send political messages, and interact with the public [24]. Also, voters often use Twitter to share their opinions on candidates, politicians, and political debate. It has been reported that a substantial percentage of voters (for example, 25 % of people in America [24]) are Twitter users, and an average of 34% of Twitter users share political opinions [25,26]. The fact that the prediction of election results based on Twitter data is quicker and cheaper than the polls, means that Twitter can be used to monitor elections in real-time [27]. Another advantage is that the sample size provided by Twitter (the number of people whose political views are analyzed) is much higher than the surveys [24]. These features have made Twitter an important source of data for election results prediction. Another benefit is that the sample size provided by Twitter (the number of people whose political views are analyzed) is much larger than that offered by the polls [24]. These features have made Twitter an important data source for forecasting political outcomes.

[1] Yalova University, Department of Industrial Engineering

[2] Muğla Sıtkı Koçman University, Department of Mathematics

[3] Muğla Sıtkı Koçman University, Department of Mathematics

A large number of studies have been conducted to forecast election outcomes via Twitter data on political patterns and elections in several different countries, such as Germany [2], Ireland [14], USA [28], [29], [30], [31], [24], Dutch [16], [32], Pakistan [33], Colombia [34], UK [35], Venezuelan [36], France [37], Indonesia [38], [9], India [27], [39], Poland [40], Spain [41], Turkey [25].

Besides, studies have been conducted to acknowledge topics that are influential in election choices [24], forecast the political attitudes of Twitter users based on their [42], [43]. Some studies aim to establish the demographics of candidates' supporters by analyzing the images of supporters of election candidates on Twitter with image processing [44].

In studies on election result prediction based on Twitter data, estimates were made using one of the factors such as the number of positive, negative, and neutral tweets posted about parties, the effect size of these tweets (the number of re-tweets), or the number of people who posted these tweets. However, no study was found that used all of these factors together. The goal of this study is to develop a new approach that takes into account all of the factors described and contributes to the literature in this context. A new model for the election result prediction based on Twitter data has been developed for this purpose. The model was tested by attempting to predict the results of the US 2020 elections in November, which had not yet taken place when the first version of this article was written. Also, a comparison has been made with alternative estimation approaches in the literature.

## 2. Related Work

Similar studies have been surveyed to develop the methodology to be used to predict election results. Data collection (tool/software, keywords, quantity), preliminary analysis (cleaning data), sentiment analysis, and election result prediction methods/tools/software used by these studies were briefly mentioned and implications were made from this information.

The study by Tumasjan et al. [2] is one of the first studies on social media data and election results prediction. It was investigated whether Twitter was used as a forum for political negotiation and whether online messages on Twitter reflect offline political sentiment. To this end, they collected tweets containing the names of the 6 parties represented in the German parliament or the names of leading politicians to predict the German federal election results. Tweets were automatically translated from German to English and Emotion analysis was performed with LIWC text analysis software [45]. As a result, it has been determined that the ranking is the same according to the number of tweets mentioning the parties and the voting share in the election results. It has been stated that Twitter is indeed widely used for political conversations, and Twitter messages reasonably reflect the offline political landscape.

Bermingham and Smeaton [14] started to make predictions by analyzing Twitter social media data before the 2011 Irish General Election and started publishing the results with the tag "# GE11 Twitter Tracker". They used both volume-based calculation and sentiment analysis in their studies where they assumed that the percentage of votes a party received was related to the volume of relevant content on social media. Before the election, they collected 32,578 tweets about the election between 8 February and 25 February using party names and abbreviations related to the five main parties, as well as the election hashtag. The tweet was analyzed using the supervised learning sentiment analysis framework they developed. In this analysis, they used tokenizers developed by Laboreiro et al. [46]

optimized for user-generated content that preserves sociolinguistic features such as emoticons (":-)") and unusual punctuation marks ("!!!!"). They then estimated votes in three different ways: the ratio of the total number of tweets about the party to all tweets; The ratio of the number of positive tweets to all positive tweets is the ratio of the number of negative tweets to all negatives. Unexpectedly, the total tweet rate was more accurate than the positive tweet rate. They stated that they correctly predicted the number of votes for 5 parties with an average absolute error rate of 1.61% only by proportioning the number of tweets related to the party to the total number of tweets.

Sang and Bos [16] collected 64,395 tweets a week before the election to estimate the 2011 Dutch election results. They filtered tweets by applying a two-step process called normalization. They eliminated the tweets containing more than one party name, and they took the first tweet of each user, that is, one tweet for analysis. After these eliminations, 28,704 tweets remain. For analysis, they grouped the tweets just based on the party name in the tweets. The authors did the sentiment analysis manually. Then, they calculated the weight of emotions by comparing the number of non-negative tweets about a party to the total number of tweets about that party. They tried to estimate the number of seats and the rate of votes won by the parties by multiplying their sentiment analysis weight by the number of tweets posted about the party before the normalization process. They claim that the parties estimated their voting rate with an absolute difference of 17.4% in total.

In the first three studies examined [2], [14], [16], it was seen that the election results were estimated by considering the number of tweets or positive tweets about the parties participating in the elections. However, in the study made by Choy et al. [28], the number of people who posted these tweets about the election was taken into account, not the number of tweets. They [28] collected 7,541,470 tweets over 3 months for the prediction of the 2012 USA Presidential Election result. They used AFINN [47] for sentiment analysis of the tweets they collected. After the sentiment analysis, they analyzed with the assumption that the person who positively tweeted about one of the presidential candidates supported that candidate. They multiplied the proportion of people who positively tweeted about a candidate by the proportion of internet users (using Twitter + not using Twitter) in the states. They also multiplied the proportion of support given to candidates in the states in the previous election, with the proportion of non-Internet users. They have calculated the total support for a candidate by adding both multiplication results. As a result of the analysis, they were able to correctly predict that Obama will win the 2012 election before the election.

Mahmood et al. [33], also took into account the number of tweeters to estimate the Pakistan 2013 election results. They identified 24 users who tweeted about the election and collected a total of 9000 tweets during the election period. They manually specified 40 words for or against the parties. According to these keywords in a tweet, they used RapidMiner grouping models to predict which party the tweet supports and which party it is against. They tried to estimate the vote rates of the parties by collecting the numbers of tweets for and against. However, they improperly predicted the election results with this small data set they gathered from very few users.

Another study that takes into account the Twitter users who tweeted about the election was made by Makazhanov et al. [42]. However, in this study, they developed a model not to predict the election results, but to predict the party that the Twitter user will vote for based on their Twitter posts. They collected tweets during the 10 days preceding the election, using 27 manually selected keywords such as party hashtags, party names, leader names, and general election hashtags related to the 2012 Albertan general election in Canada. As a result, they collected 181,972 tweets from 28,087 accounts. They removed the party candidates' accounts and those with a communication language other than English from these accounts and identified 24.507 accounts. Distant supervision approach was used for the analyzes. Tweets by 252 candidates of the four major parties that entered the election were used for the training of the model. They developed a model that classifies Twitter users as a party's supporters based on their characteristics such as retweeting, following, and interaction with the party. They also tested the model they developed for the 2013 Pakistani general elections. As a result, they stated that the model they developed could be used to predict the political preferences of Twitter users.

Ramteke et al. [29] again tried to estimate the election result based on the number of tweets. They collected tweets for Donald Trump and Hillary Clinton, two candidates who participated in the 2016 USA Election, on March 16-17, 2016, containing the names of the parties and candidates. They used 60 thousand tweets they obtained for analysis. They analyzed the tweets containing the hashtags used by the supporters of a party/candidate with the assumption that it could be positively tagged for that candidate/party. With this assumption, hashtags used at high frequency in the Tweet dataset (minimum 20 in this study) were manually tagged as positive or negative. For example, the #MakeAmericaGreatAgain hash has been labeled as positive support for Trump. In their analysis with the Phyton software, they first labeled the emotional polarity of the sentences with the VADER algorithm [48]. Then, by applying the Multinomial Naive Bayes and Support Vector machines machine learning models with the Python Scikit-learn package, they classified the tweets as positive and negative for the candidates. The estimation of the election result by dividing the total number of positive tweets sent about the candidate by the total number of tweets related to that candidate turned out to be incorrect.

In the thesis study by Guzmánn [34], tweet data were used to predict the 2014 Colombia election results. In the study, a machine learning approach was used to detect Spam tweet accounts. They emphasized that it is necessary to detect and weed out spam accounts for election result prediction analysis based on Twitter data, and to take into account the texts produced by users when performing sentiment analysis for tweets. As a result of the study, they stated that social media analysis has the potential to predict the voting rate of parties and to rank the highest-rated candidates accurately.

Burnap et al. [35] collected tweets containing the names of the parties or candidates for more than 3 months to estimate the 2015 UK General Election results. They applied sentiment analysis to around 13 million tweets they obtained using the software developed by [49]. They removed the tweets with content related to more than one party/candidate from the data set. They tried to estimate the vote ratio by collecting and rating all positive tweets about the party/leader. As a result of the study, although the parties incorrectly predicted the number of seats in the parliament, they correctly determined the voting ratio order.

Wicaksono et al. [30] tried to predict the election results of states by collecting English tweets containing any keyword (e.g. republican, democrat, Hillary Clinton, Donald Trump) and location information about parties or candidates participating in the 2016 US presidential election. Using Sentiment140 tweet corpus [50] that contain 1,600,000 data sets (800,000 training data for positive and negative emotion) and 497 test data (181 positives, 177 negatives, and 139 neutral) and abbreviations dictionary [51], they performed sentiment analysis of tweets by 3 types of sentiment classifiers (Binarized Multinomial Naïve Bayes Classifier, SentiWordNet, and AFINN-111). As a result of the analysis, they considered the tweet about a party containing positive emotion as a positive vote for that party, and the one containing negative sentiment as a positive vote for the opposite party. This approach probably led to the erroneous calculation that the same person would vote positive votes for different parties due to different tweets. As a result of the study, it was wrongly predicted that Hillary Clinton (Democrat) won the election with a total of 253 electoral votes, as Donald Trump (Republican) received only 219 electoral votes.

Wicaksono [30] and Ramteke [29] shows that it may not be sufficient to use only the number of total or positive tweets about parties to predict the election outcome, on the contrary to the first studies [2], [14], [16]. As a matter of fact, with the advancement of techniques such as artificial intelligence and machine learning, more advanced methods have been used for both sentiment analysis and voting rate estimation in election results prediction studies based on social media data. Another example of this kind of study is the work of Castro et al. [36] to predict the results of the 2015 Venezuelan Parliament election. In this study, they analyzed 60 thousand tweets published within the geographical borders of Venezuela one week before the election. In the study where they used social network analysis and unsupervised machine learning techniques, they were grouped according to states by using latitude and longitude information in tweets. They made predictions according to the frequency of the words they determined about the election in the tweets. Although they could only estimate the election results at a rate of 79.17%, they stated that they guessed the winning party correctly for 21 of the 24 states (87.50%) in Venezuela.

Anuta et al. [31] conducted a study comparing the election forecast model based on Twitter data with polls. They used previously collected tweet data about the 2016 US elections in the study. They filtered the tweets in the data set according to the location (USA) and the language of the Twitter user. They saved 3 million tweets of 750,000 users to the PostgreSQL database for analysis. VADER [48] application was used for sentiment analysis. According to the tweets of the people about the candidates, they decided whom they would vote for and estimated the votes by proportioning the number of people. As a result of the study, they stated that Twitter was more biased than surveys and the prediction was more erroneous with data based on tweets. However, this may be due to errors in their prediction methods based on Twitter data they use. As a matter of fact, if two candidates are mentioned in a tweet, one candidate may be praised and there may be a negative judgment against the other candidate. It is wrong to evaluate this tweet positively and analyze it as positive vote for both candidates.

In the study conducted by Wan and Gan [37], unlike previous studies, they estimated the election result taking into account not only the positive or negative but also the number of neutral tweets about the parties. First of all, they collected tweets about the 2017 French elections that were posted before the election and included the names of the candidates. They identified words in the tweets that indicate a positive or negative Emotion for a particular candidate. For example, they stated that the word Obama was positive for Macron, one of the candidates, because Obama published a video supporting Macron. They categorized these words for and against candidates. They categorized the tweets as positive or negative for candidates based on the emotion-expressing words in their content. Tweets that do not express emotion are categorized as neutral. They used the number of tweets to predict the election result. They proportioned the total number of positive tweets about the candidate to the number of positive and negative tweets about the candidate and multiplied by the ratio of all tweets (including neutral) about the candidate among all tweets in the sample. They estimated the daily election results according to the tweets they collected using the formulas they specified. On the last day, they predicted the election results with only a 2% margin of error. They stated that their vote rate estimates, taking into account the neutral tweets, were more successful than the methods considered only for positive and negative tweets.

Toker et al. [43], similar to the study of Makazhanov et al. [42], tried to determine the political tendency of Twitter users. They stated that, according to the words used in the tweets of Twitter users, they determined which party supporters (political tendencies) they were. For their analysis, they selected 1011 Twitter users by random sampling from among the followers of a comedian who used their real name and real picture and who posted at least 130 tweets in the last three months. Depending on their tweets or retweets that contain words related to Turkey's 2015 June elections,

These users' political tendencies were tried to be identified. The Twitter messages of 364 people selected among these users were re-analyzed for the 2015 November elections and investigated the changes in the political tendency. They stated that the political tendency of 68.2% of the sample they determined was determined.

Grover et al. [24] collected a total of 784,153 tweets about the election in three months in their study after the 2016 USA election. Tweets were collected using keywords related to candidates and parties. Besides, all the tweets of the two candidates of the election (Hillary Clinton "and" Donald Trump ") were collected. Tweet collecting process and sentiment analysis were done in the R programming language with syuzhet, lubridate, and dplyr libraries. For Topic modeling, tm and topicmodels libraries of R were used. According to this study, discussions on Twitter can polarize users, that is, they affect them. They analyzed the polarization using the Newman model [52]. As a result of the geographic analysis of the tweets, they also found that Twitter causes ideas and ideologies to affect other users (acculturation). Acculturation has been defined as the change in preferences within an individual when interacting with individuals or groups with a different cultural background [53]. Another interesting result of the study is that although the net positive (positive-negative) tweets about Clinton are higher than Trump's, the average number of likes for Trump's tweets is almost 3.8 times that of Clinton. Now that Trump won the election, the conclusion to be drawn from this study is that it would be useful to include not only the number of tweets but also the number of likes and retweets in the analysis.

Another study that takes into account the number of people who tweeted about the election is the study of Bansal & Srivastava [27]. Using the R programming language, they had collected more than 300,000 tweets related to India Uttar Pradesh legislative elections. They used words including party names, party leader names, election campaign slogans, and hashtags as keywords for collecting related tweets. They performed sentiment analysis on these tweets with a method they developed themselves and called Hybrid Topic-Based Sentiment Analysis (HTBSA). With this method, a sentiment score was calculated for each tweet. They estimated the election result by taking into account the total number of Twitter users who had sent positive tweets about the parties. The fundamental assumption in this calculation is that a person's positive tweet about a party is the intention to vote for that party. As a result of this study that had conducted after the election, they reported that they predicted the election results with an average absolute error of 8.4%. They also claimed that the HTBSA method they developed was more successful than other lexicon-based sentiment analysis methods.

Even after 8 years have passed since the first studies [2], there are still studies that take into account only positive tweet counts. For example, in the study conducted by Budiharto & Meiliana [9], using the R programming language, tweets containing the determined hashtag (#) related to the Indonesian 2018 selection were collected. Sentiment analysis was applied to the pre-processed tweets using the TextBlob library. They predicted with a simple approach that the candidate with a high number of positive tweets will win the election. They stated that their prediction turned out to be correct as a result of their choice.

In the social media application [39], which was made to predict the Indian Punjab state election results, data was taken from Twitter using hashtags related to the parties. Only 9157 tweets were collected and all tweets were translated into English. Before the analysis, punctuation marks, numbers, web page links, extra spaces were removed from the content of the tweets as a pre-process. Tweets are associated with parties by hashtag (#) or account name (@) found in tweets. Network analysis was performed to visualize how many tweets were posted about which party. They performed sentiment analysis on the tweets. Then they subtracted the total number of negative tweets from the total number of positive tweets about a party and proportioned this figure by subtracting the total number of negative tweets from the total number of positive tweets of all parties. They accepted the rate they found as the voting rate. From this ratio, they used simple linear regression to estimate the number of seats in the council. They state that the estimation of the number of seats they obtained is correct as a result of their analysis using the voting rates and the number of seats of previous years.

In the study by Grimaldi et al. [41] sentiment analysis and related tweet amounts were used to predict the 2019 Spain election results. One million tweets covering parties, candidates, and election-related hashtags have been collected from Twitter. Before analysis, they deleted unwanted phrases, words, URLs, numbers, and dates from these tweets. Using the frequencies of the expressions in the tweets, they extracted the attributes with the R software. They used five different machine learning algorithms in their analysis and compared their performance. They obtained the most accurate selection result estimation with the kNN (k-Nearest Neighbor) classification algorithm.

In the study of Sanders and Bosch [32], it was investigated which type of tweets (making a positive comment, recommending a candidate, etc.) are more effective in predicting the election result. For this purpose, 17,000 tweets were manually tagged by 500 people at least three times for 10 different categories. In other words, each tweet is tagged for 10 different categories such as mockery, recommendation, positive or negative according to the word it contains. Then, using these 10 categories (annotations), the election results were tried to be predicted. It was analyzed which of these 10 categories correctly predicted the election results with the lowest margin of error. This study was conducted for three different elections in the Netherlands. As a result, a single category with the lowest error could not be determined among all three selections. Therefore, the most useful category of tweet suggestions (eg positive, suggestive, etc.) could not be submitted for the prediction of election results.

In a study conducted for the elections in Turkey [25], changes in the interest of Twitter users to the parties have been estimated depending on the number of daily tweets. During, the general election in 2015, the referendum in 2017, and the general election in 2018, tweets containing the names of the parties or election slogans or the names of their leaders were collected. More than 3 million tweets were collected during each election period. In the analysis, they only calculated the daily changes in the number of tweets. But they did not apply sentiment analysis to tweets. They group the tweets as positive or negative according to the hashtag/keyword they contain. They calculated the interest of Twitter users by looking at the number of tweets posted. As a result of the study, they stated that they could detect the change of Twitter users' interest in parties on daily basis.

In the study of Chauhan et al. [23], previous works made for the prediction of the election result using social media were examined. Data collection and analysis methods used in these studies were compiled. Suggestions have been made for researchers to conduct similar studies. Within the scope of the article, 38 articles related to the election result estimation based on social media data published between 2010 and 2019 were examined. It was seen that in 23 of these articles, data were collected from Twitter social media. It was mentioned that sentiment analysis and volumetric approach were used as the analysis method for the selection outcome prediction. It was stated that in 24 of the studies, the election results were correctly estimated. While most of the studies (29 of them) were conducted after the election, only a few (9 of them) were carried out before the election. It is emphasized that it is more meaningful to state the election results before the elections. It was reported that to obtain a more accurate election result prediction, attention should be paid to the geographical location and age limit to determine and eliminate non-voters' tweets. It was also emphasized that it would be beneficial to remove spam or mocking messages and tweets belonging to bot accounts. It was pointed out that analyzing different social media data such as Facebook and Instagram in addition to Twitter will increase the prediction performance. As an analysis method, it was mentioned that deep learning gives more accurate results.

As a result of the literature review, the factors that are found to be effective in the election result prediction based on social media data are the number of positive, negative, and neutral tweets, the effect size of these tweets (the number of re-tweets), and the number of people who posted these tweets. It is thought that a new method that takes all these factors into account will give more effective results than existing methods and contribute to the literature in this context. Below, a new prediction model based on Twitter data has been developed for this purpose. The model has been tested by trying to predict the result of the USA 2020 November elections, which had not yet occurred at the time of publishing the first version of this article.

## 3. Methodology

According to the previous studies, in social media analytics applications, the number of people tweeting for/against a party; the number of positive, negative, and even neutral tweets about a party; and the amount of retweeting of these tweets are important factors for predicting election results. A prediction model that takes all these factors into account could not be found in the literature. So, in this study, a new election result prediction model based on social media data was created, which takes all the aforementioned factors into account.

Election results prediction studies are generally carried out after the election, but publishing them before the election is more meaningful [23]. Therefore, to evaluate the success, the newly developed model was used to predict the result of the general election of the USA in November 2020 which has not yet been realized at the time of publishing the first version of this article.

The selection result estimation obtained with the newly developed model has been compared with other estimation models in the literature and survey results. The compared models are the prediction model based on the number of tweeter users [27], [31], [28]; prediction model based on the number of positive and negative tweets [39]; also prediction model based on positive, negative, and neutral tweet counts [37].

### 3.1. Data Collection

In the examined studies, it was observed that the keywords used in collecting the election-related tweets included the names of the parties, their abbreviations, the names of the party candidates, and the election slogans. So, in this study, data were collected using similar concepts. Tweets containing at least one of the 29 keywords about the four parties that received the most votes in the last election, were collected. These keywords are given in Table 1. Tweets have been collected since July 1, 2020. It is planned to continue the data collection process until November 11, 2020, ten days after the election.

The number of tweets collected is more than 20 million. The election results estimation study made using more than this amount of tweets in the literature could only be identified in 3 [18,35,54]. Therefore, the number of tweets collected is considered to be sufficient for analysis.

In some of the reviewed studies, the election-related tweets were collected in the last month or less before the election [27], [36], [16], [14], and even in some studies [29], it was seen that they gathered just a day before the election. In this study, data were collected four months before the election. This long time allows a better analysis of the changing popularity of parties.

The collected tweets are divided into four different groups as the first 40 days (from July 1 to August 12, 2020), the next 40 days (from August 13 to September 25), the last 40 days (from September 26 to November 3) and the first 10 days after the elections (from November 4 to 14). The analyzes were applied to these groups separately. By comparing the results of the analysis, the changes in the period before and after the election were also analyzed.

*Table 1 Keywords*

| Category | Keyword(s) |
|---|---|
| US November 2020 Election | #USAelection OR #NovemberElection |
| Democratic Party | @DNC OR @TheDemocrats OR Biden OR @JoeBiden OR "Our best days still lie ahead" OR "No Malarkey!" |
| Green Party | @GreenPartyUS OR @TheGreenParty OR "Howie Hawkins" OR @HowieHawkins OR "Angela Walker" OR @AngelaNWalker |
| Libertarian Party | @LPNational OR "Jo Jorgensen" OR @Jorgensen4POTUS OR "Spike Cohen" OR @RealSpikeCohen |
| Republican Party | #MAGA2020 OR @GOP OR Trump OR @POTUS OR @realDonaldTrump OR Pence OR @Mike_Pence OR @VP OR "Keep America Great" |

### 3.2. Sentiment Analysis

Sentiment analysis is an automatic text processing technique used to measure views and attitudes. Although widely used in social media data to assess general perceptions of products and brands, it can provide insight into events and issues. Rather than relying on manual techniques to identify positive or negative themes, sentiment analysis compares terms to a vocabulary dictionary with predetermined emotion scores. In addition to the wide variety of methods utilizing different dictionaries and scales, the standard "Syuzhet" R package is used in this study.

The package comes with four sentiment dictionaries and provides a method for accessing the robust, but computationally expensive, sentiment extraction tool developed in the NLP group at Stanford [56]. The default "Syuzhet" lexicon was developed in the Nebraska Literary Lab under the direction of Matthew L. Jockers is used in this study. Within each tweet, each word is evaluated independently and a score is given, and a total score is returned for each tweet. Positive scores indicate positive attitudes, negative points indicate negative attitudes, and words not found in the corresponding dictionary get zero.

### 3.3. Multifactor Twitter Data Analysis Method

Our most important contribution in this study is to analyze by giving weight not only to the number of views of Twitter users but also to the views of the most influential users in the network structure formed by users speaking about US Elections 2020.

For the network creation process, we first determine the users who have interaction among them throughout the mentioning or direct messages. Such interactions are extracted from the data set and a weighted directed network is formed. Each user has a fixed User-Id; hence, we name each vertex after the User-Ids. The weight function is chosen to be the number of Retweets each interaction has.

The most influential users in such a network corresponding to the vertices throughout their PageRank index. PageRank centrality in a network is a measure that can be summarized as; a vertex is important if it is linked from other important and link parsimonious vertices if it is highly linked [57]. Three important factors determine the PageRank of a vertex. The first factor is the number of edges the vertex receives. This factor emerges from the idea of the more edges a vertex attracts, the more it is perceived. The second factor is the edge propensity of the vertices which depends on the idea of the number of edges coming from parsimonious vertices which are more valuable than those emanated by spendthrift ones.

Let $A = (a_{ij})$ be the adjacency matrix of a directed network. Then, the PageRank centrality $x_i$ of the user, $i$ can be defined by

$$x_i = \alpha \sum_k \frac{a_{k,i}}{d_k} x_k + \beta,$$

where $\alpha$ is a damping factor, $\beta$ is an exogenous variable, $d_k$ is out-degree of vertex $k$. Then, in matrix form we have

$$x = \alpha x D^{-1} A + \beta,$$

where $\beta$ is a positive constant vector which is called the personalization vector, $D^{-1}$ is a diagonal matrix whose $i$-the diagonal element equals to $1/d_i$. For large networks like the Twitter network, we prefer to use the power method for the computation of PageRank [58, 59]. Moreover, the damping factor $\alpha$ should be chosen between 0 and $1/\rho$, where $\rho$ is the largest eigenvalue of the matrix $D^{-1}A$, to guarantee the convergence of the measure.

In the network of users speaking about American elections, the number of retweets of tweets as well as measurements of users' activity is an indicator of participating in positive or negative opinions [60]. Considering the emotion score of a tweet posted by a user, the centrality of the user in the interaction network, and the number of retweets of the tweet, a score as follows is determined for each user in this study:

$$MFS_i = 1000 x_i \sum S_j R_j,$$

where $x_i$ is the PageRank centrality of user $i$, $S_j$ and $R_j$ are the sentiment score and retweet number of the $j$-th tweet by user $i$, respectively, and 1000 is a normalization coefficient. In the data set, the retweet numbers of direct interaction tweets are found to be mostly 0. Hence, we consider $R_j = 1$ for these cases.

Unlike the methods in the literature, the autoregressive fractionally integrated moving average model (FARIMA) is used in the study for the estimation of the daily average $MFS_i$ scores on 3rd November 2020 in the datasets that we grouped as Democrats and Republicans between 1st September 2020 and 2nd November 2020, instead of the $MFS_i$ totals. Thus, it is expected to predict the election outcome depending on the multifactorial development of the discourse for both groups.

FARIMA models generalize ARIMA (autoregressive integrated moving average) models by allowing non-integer values of the differencing parameter $d$. Formally, let $\{X_t\}$ be a stationary process such that

$$\phi(B)(1-B)^d X_t = \theta(B) Z_t, \quad \{Z_t\} \sim WN(0, \sigma^2),$$

where

$$(1-B)^d = \sum_{k=0}^{\infty} \binom{d}{k} (-1)^k B^k$$

and $|d| < 0.5$. Hence, by considering $\langle MFS_t \rangle \sim X_t$ at day $t$ between 1st September 2020 and 2nd November 2020, we can forecast the $\langle MFS_t \rangle$ on 3rd November 2020 with convenient parameter estimations.

# 4. Findings

## 4.1. Network Analysis

On Twitter, while users tweeting about their views, they are also able to interact with each other through the mentioning. In order for the user to send a direct message to any other user, they must put an "@" sign in their tweet. With this interaction, it is possible to create a social network of users talking about the USA 2020 Election. In the data set, each user has a single ID number. This ID number is a positive integer. While creating the data set, if the tweet sent by a user is not a direct message to any user, the "$From\ User\ ID$" value is taken as the user's ID number and "$To\ User\ ID$" value is taken as $-1$. If the user's tweet is sent as a direct message to another user, the "$From\ User\ ID$" value is taken as the ID number of the user and the "$To\ User\ ID$" value, the other user's ID value. That is, any directed link representing an interaction is encoded with the tuple $(From\ User\ Id, To\ User\ ID)$.

Considering the possibility that users with different views may interact with each other, no user was labeled as Republican or Democrat, and the interaction network was built entirely through messaging. Although it is possible for a user other than two interacting users to retweet the reply, this situation was observed very rarely in the data set. For this reason, edge weights are not taken into consideration. In addition, the network structure was taken as directional, since any interaction does not have to cause any counter-interaction.

Topological measurements of the weightless-directional network of users speaking about USA 2020 Elections inform us about the characterizations of interaction amongst users. In this study, the well-known topological measurements Average Path Length, Global Clustering Coefficient, and Vertex Betweenness are used.

The average path length is a concept defined as the average number of steps along the shortest paths for all possible network node pairs in the network topology. It is a measure of information or mass transport efficiency in a network [61,62] and can be computed as follows:

$$Average\ Path\ Length = \frac{1}{n(n-1)} \sum_{i \neq j} d_G(i,j),$$

where $n$ is the number of vertices and $d_G$ is the distance on the directed network $G$.

Global clustering coefficient is a network clustering measure of the degree to which vertices in a network tend to cluster together and is based on 3-cliques in a network [63,64]. This measure gives a global indication of the clustering in the whole network and can be successfully applied to undirected networks as follows:

$$Global\ Clustering\ Coefficient = \frac{3 \times number\ of\ triangles}{number\ of\ all\ triplets},$$

where a triplet is three vertices that are connected by either two (open triplet) or three (closed triplet) edges and a triangle is a subgraph that includes three closed triplets, one centered on each of the vertices.

The vertex betweenness is defined by the number of shortest paths going through a vertex. Vertex betweenness is directly related to a network's connectivity, in so much as high betweenness vertices have the potential to disconnect graphs if removed [65]. The vertex betweenness of a vertex $v$ can be expressed as follows:

$$Vertex\ Betweeness(v) = \sum_{u \neq v \neq w} \frac{\gamma_{uw}(v)}{\gamma_{uw}},$$

where $\gamma_{uw}$ is the total number of shortest paths from vertex $u$ to vertex $w$ and $\gamma_{uw}(v)$ is the number of those paths that pass-through $v$.

In Figure 1, we present resulting network measure values of average path length, global clustering coefficient, and mean vertex betweenness of directed networks emerged from the interaction of the users talking about the USA 2020 election.

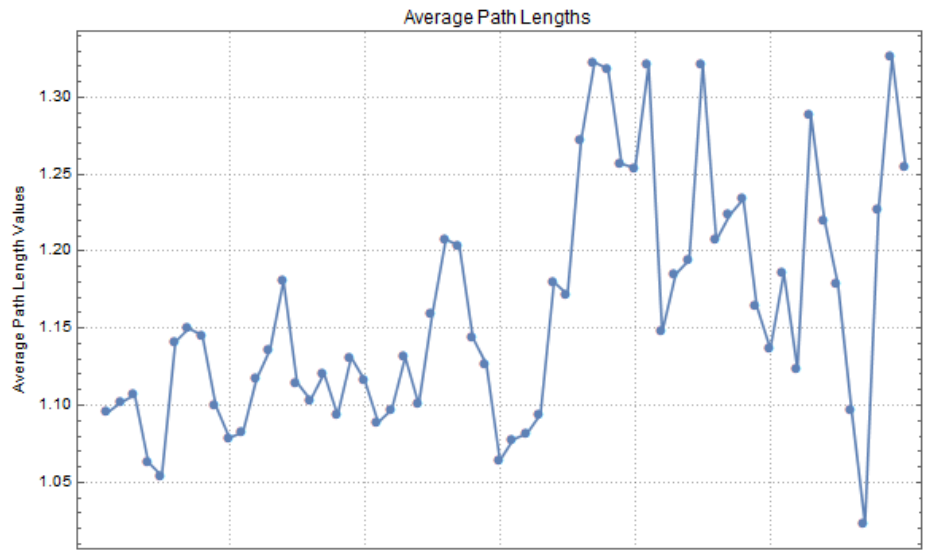

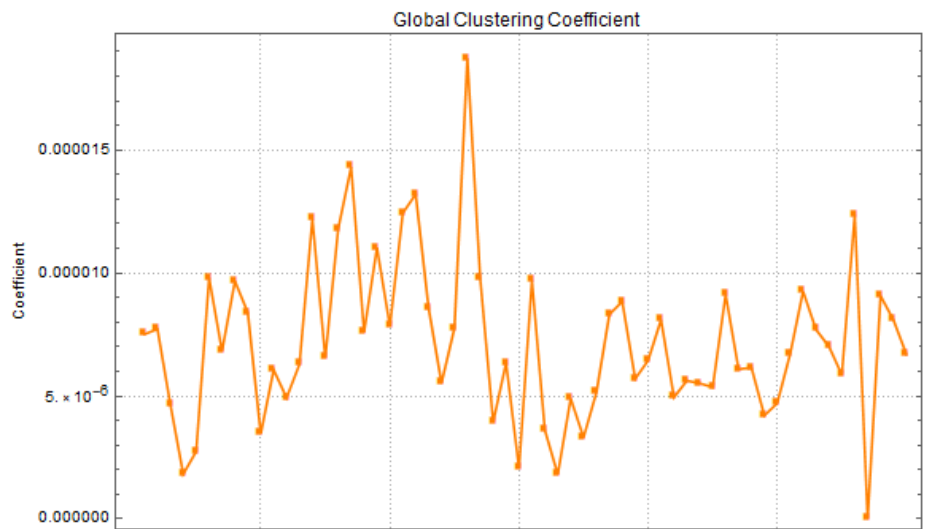

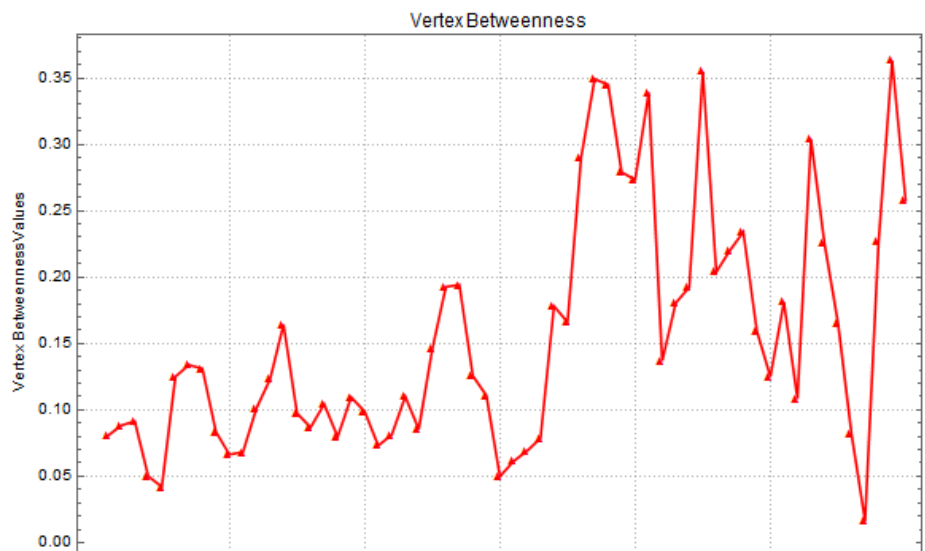

Figure 1 Network measure values

## 4.2. Descriptive Statistical Analysis

The datasets are grouped as Democrats and Republicans from September 1, 2020, to November 2, 2020. A total of 3851293 tweets were posted for Democrats and 7109941 for Republicans. Among these total tweets, the positive tweet rate for Democrats is 43.19%, the negative tweet rate is 42.57%, the neutral tweet rate is 14.24%, for the Republicans the positive tweet rate is 39.82%, the negative tweet rate is 53.33%, and the neutral tweet rate is 6.85%. See Figure 2.

Users who speak English about the elections on Twitter have the 100 highest PageRank score in the interaction network have statics as follows: for Democrats, the rate of positive tweets is 43.18%, the rate of negative tweets 41.97%, the rate of neutral tweets 14.85%, for Republicans the rate of positive tweets 37.85%, and the rate of negative tweets 50.46%, neutral tweet rate 11.69%. Besides, it has been observed that active users who speak for Democrats and Republicans use strong rhetoric, especially when speaking positive or negative about Republicans. See Figure 3.

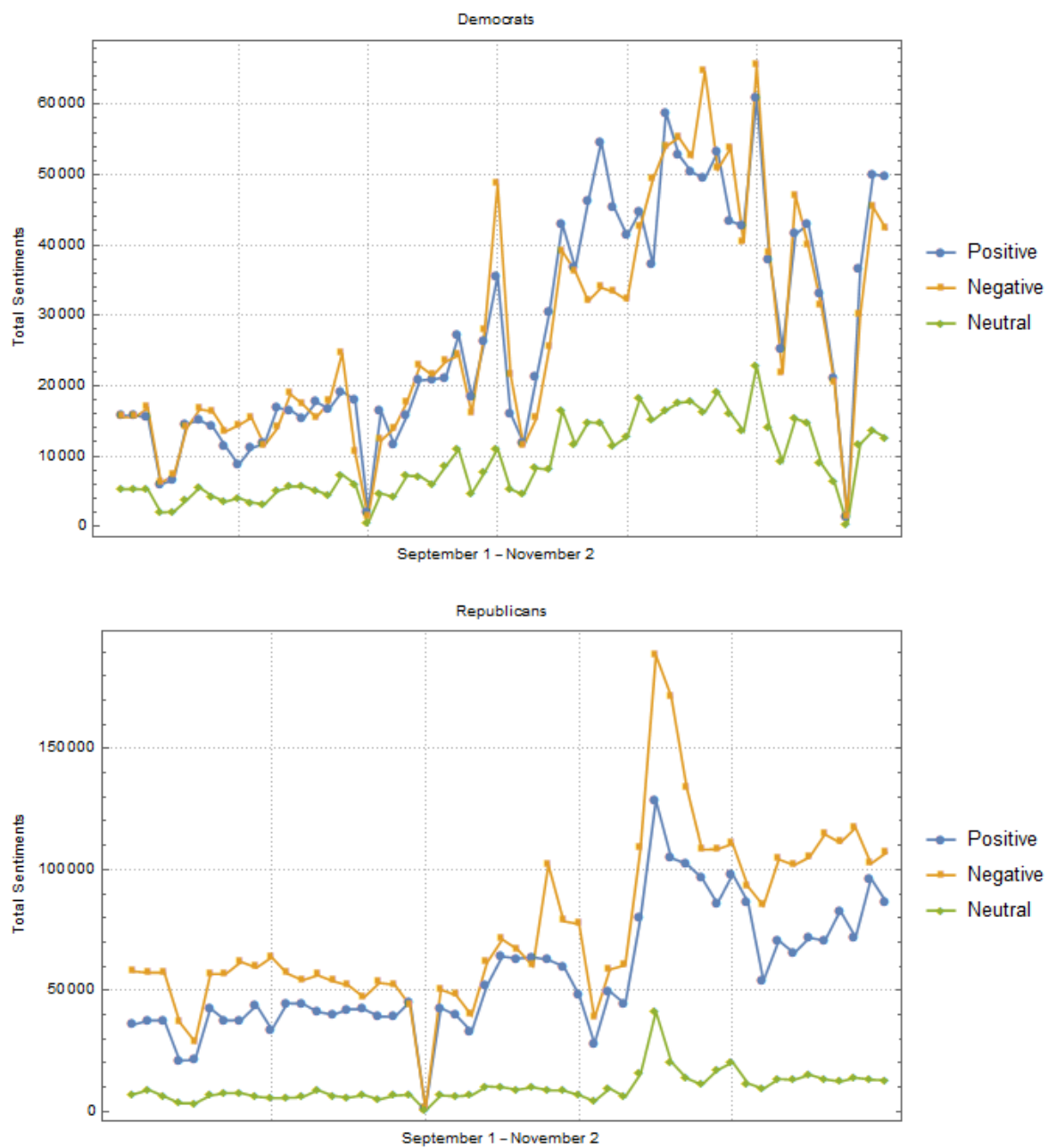

*Figure 2 The numbers of positive, negative, and neutral tweets*

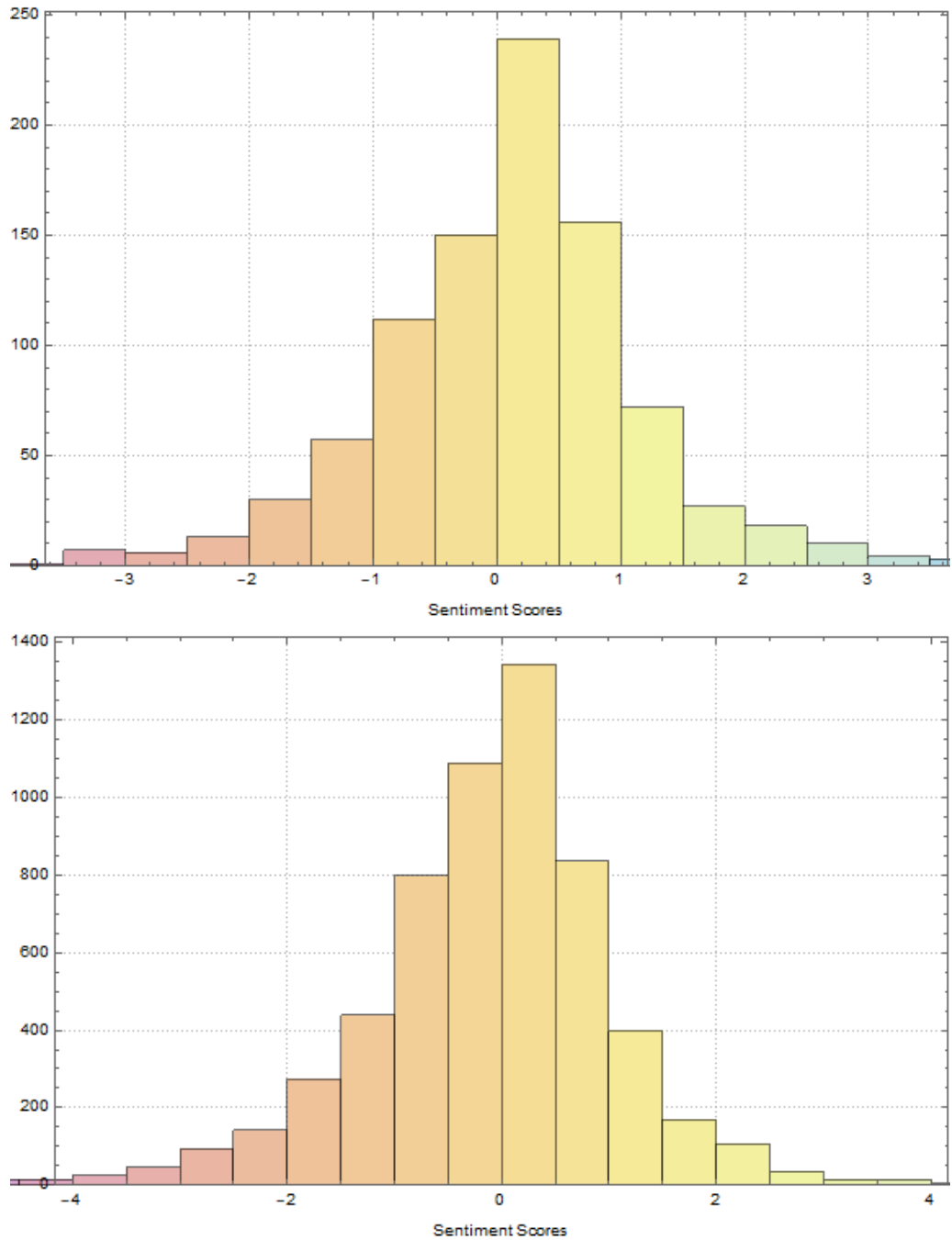

Figure 3 Scored sentiments of the users with the first hundred highest PageRank centrality

## 4.3. USA November 2020 Election Results Prediction Analysis

$MFS$ score we present in this study covers the multifactor such as the centrality of the user in an interaction network and the scored sentiments. Hence, we can obtain a number for each user in the analysis. These values vary from user to user, therefore, for each day $t$ we also obtain a distribution of $MFS_t$. The empirical distributions mostly fit Student T distribution. In Figure 4, we present Wasserstein-1 distances of each distribution.

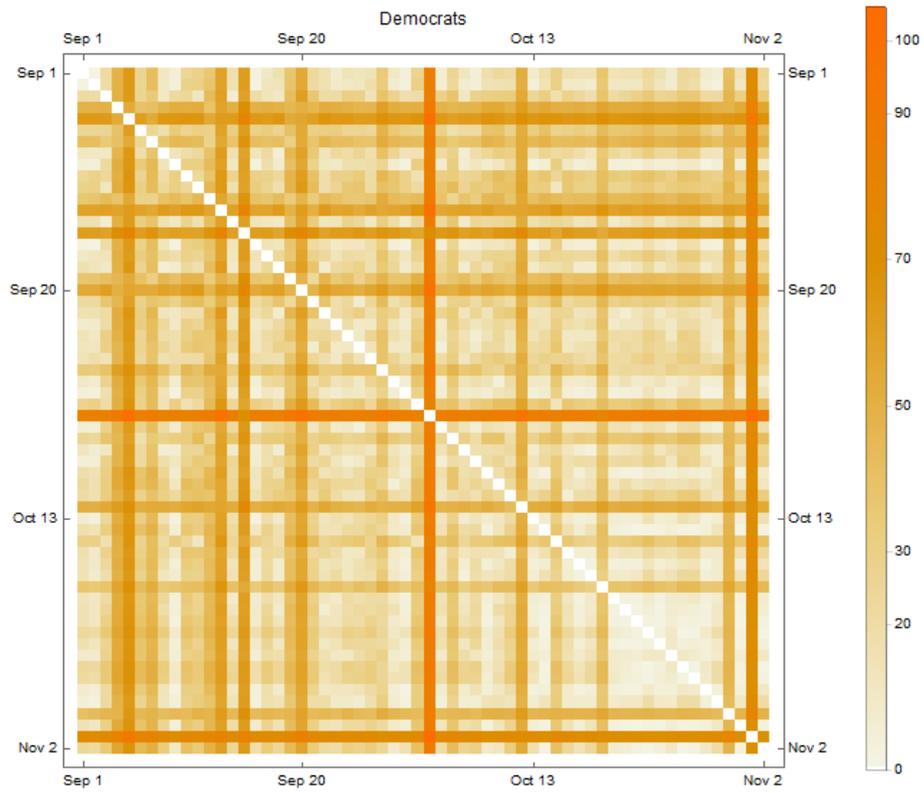

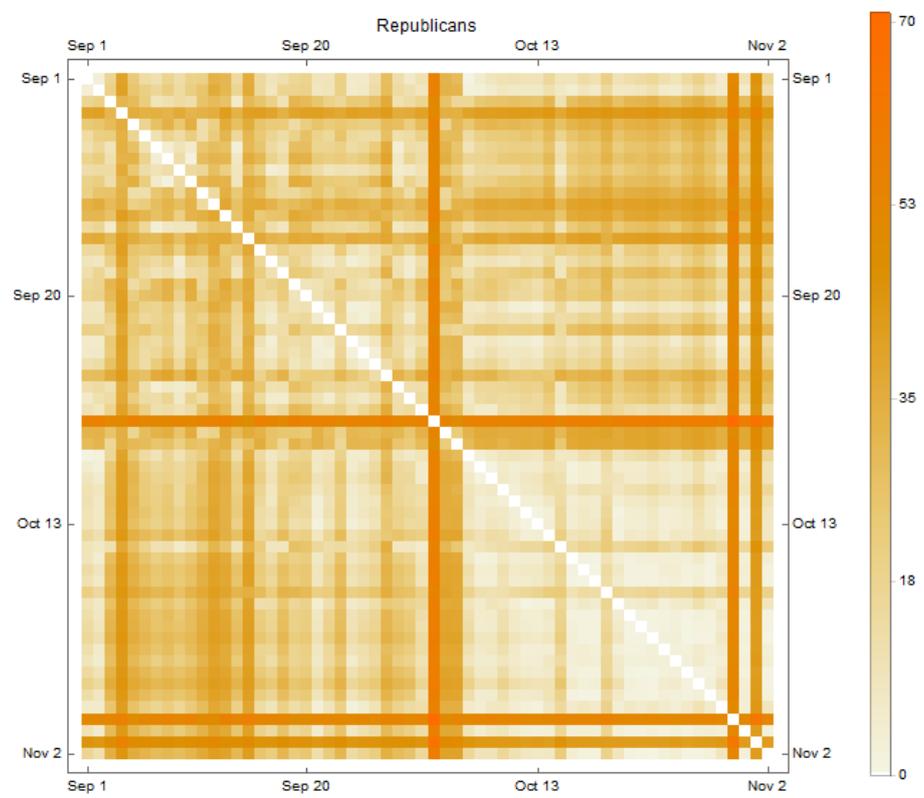

*Figure 4 Distance matrices of MFS distributions respect to Wasserstein-1 distance*

Our forecasting method relies on the distributions of each $MFS_t$. Therefore, we obtain time series for both Democrats and Republicans between September 1st – November 2nd involving the medians of each distribution denoted by $\langle MFS_t \rangle$. The fitted FARIMA model forecasts for 3rd November read as 0.004256 for Democrats and -0.010304 for Republicans. In Figure 5, we present the time series and forecasts.

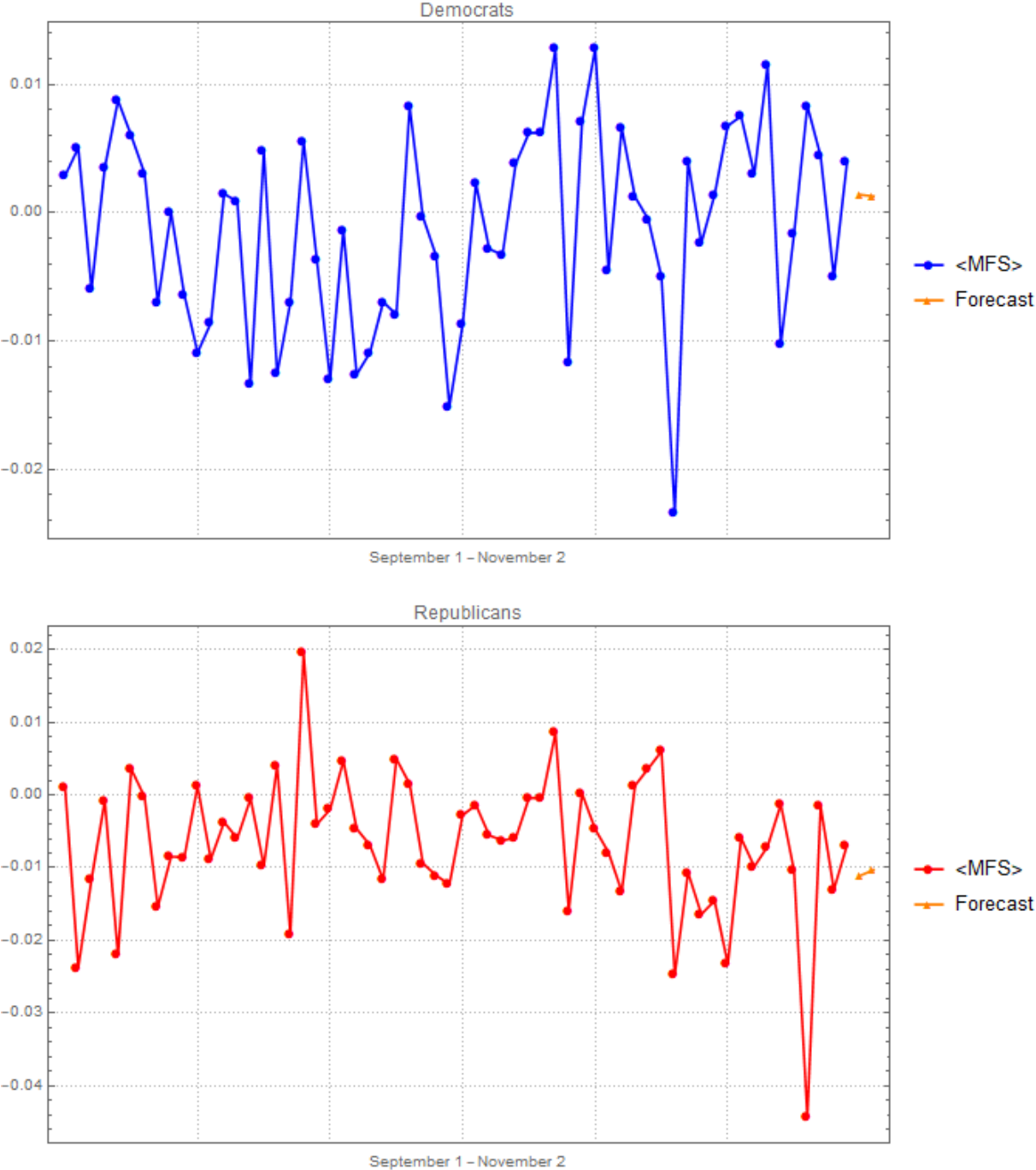

*Figure 5 Forecasts of time series $\langle MFS \rangle$ by FARIMA*

In order to analyze how FARIMA forecasting changes on last month, we give FARIMA forecast values on November 3rd of $\langle MFS \rangle$ starting from October 5th in Figure 6.

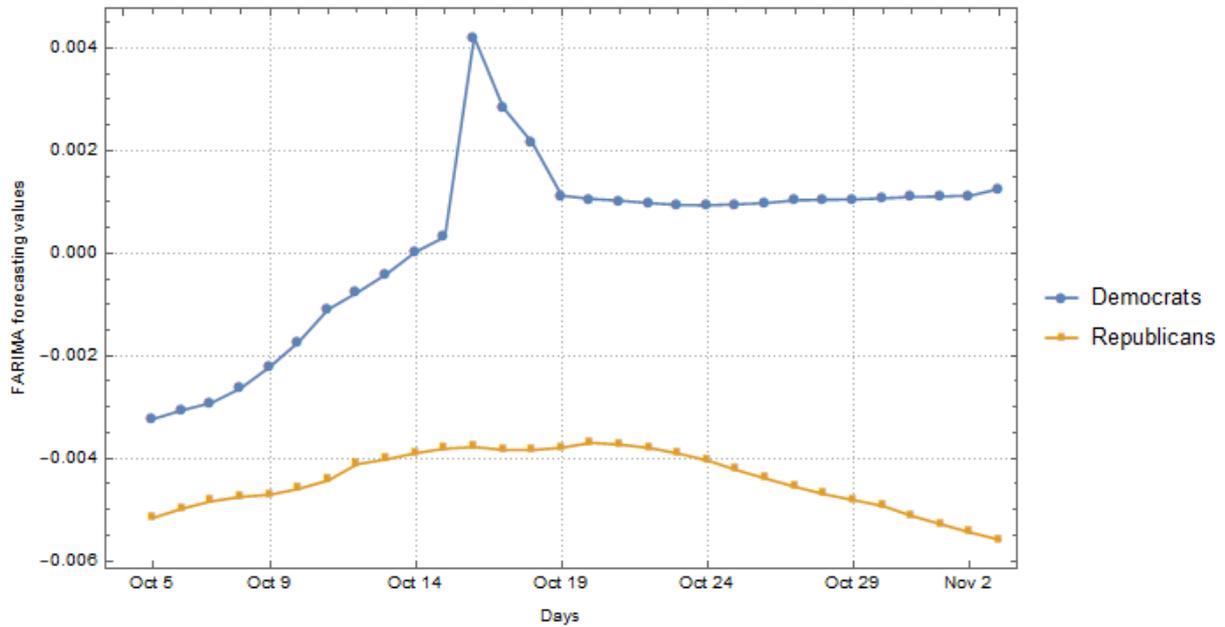

*Figure 5 FARIMA forecasting values starting by October 5th*

The fitted FARIMA model forecasts are used to predict the trends of (MFS) values which lead us to have the vote share. Assuming that these scores represent the difference in voting rates, the difference in voting rates between the two parties was estimated as 1.45% (0.004256 +0.010304). And with the assumption that both parties will get all the votes, the voting rates estimated by the multi-factor method according to this difference are determined as 50,73% for Democrats and 49,27% for Republicans.

## 5. Comparison of Alternative Methods

In order to see the success of the developed multi-factor method, the estimation values obtained by this method were compared with other methods selected from the literature [30], [37], [39], and the polls [55].

The forecasts of the election results made by The Economist [55] using 464 polls conducted on different dates within the scope of 85 different polls research were considered as the results of the polls.

In the first alternative method [39] used for comparison, only positive and negative tweet counts were taken into account. In this method, in order to estimate the vote shares, the percentages called actual sentiment score calculated by the following formula are used for the estimation of vote shares:

$$Actual\ Sentiment\ Score\ (Vote\ Share\ of\ Party\ A)\% = \frac{\text{pos}(A) - \text{neg}(A)}{T(A) + T(B)}$$

Where pos(A) and neg(A) are the Total Number of positive and negative tweets for party A respectively. T(A) and T(B) are the total numbers of tweets that are related to candidate/party a and candidate b, respectively, including positive, negative, and natural. However, as seen in Table 3, applying this formula to the number of tweets collected in this study produced meaningless results.

In the second alternative method [37], we chose a formula that takes into account the number of neutral tweets in addition to the positive and negative number of tweets:

$$popularity\ (vote\ share\ of\ party\ (A))\% = \left[\frac{pos(A)}{pos(A) + neg(A)}\right]\left[\frac{T(A)}{T(A) + T(B)}\right]$$

Again, the results obtained with this formula did not make much sense.

Finally, a formula developed by using the approach suggested by Wicaksono et al. [30]. According to Wicaksono et al. [30], a positive tweet about a candidate indicates the intention to vote for that candidate, and a negative tweet indicates the intention to vote for an opposition candidate. Using this approach, as the third alternative method, vote share prediction is formulated as follow:

$$(vote\ share\ of\ party\ (A))\% = \frac{pos(A) + neg(B)}{T(A) + T(B)}$$

Eventually, a result close to the actual election results was obtained with this formula. However, the result obtained with this formula was not as successful as the polls and the multi-factor method. As can be seen in Table 4, the multi-factor method is the most successful method with a lower average absolute error rate than all 3 alternative methods results and polls.

*Table 2 Positive, Negative, and Neutral Tweet Numbers*

|  | Total Tweet | Positive | Negative | Neutral |
|---|---|---|---|---|
| Democrats | 3.851.293 | 1.663.373 | 1.639.495 | 548.424 |
| Republicans | 7.109.941 | 2.831.179 | 3.791.732 | 487.031 |
| Total | 10.961.234 | 4.494.552 | 5.431.227 | 1.035.455 |

*Table 3 Result of Alternative Methods and Polls*

|  | Multifactor Model | Polls [55] | First alternative method [39] | Second alternative method [37] | Third alternative method [30] | Real Results |
|---|---|---|---|---|---|---|
| Democrats | 50,73% | 54,40% | -2,55% | 17,69% | 54,96% | 51,40% |
| Republicans | 49,27% | 45,60% | 102,55% | 27,73% | 45,04% | 46,90% |
| Total | 100,00% | 100,00% | 100,00% | 45,42% | 100,00% | 98,30% |

*Table 4 MAE of Alternative Methods and Polls*

|  | Multifactor Model | Polls [55] | First alternative method [39] | Second alternative method [37] | Third alternative method [30] |
|---|---|---|---|---|---|
| Democrats | 0,67% | 3,00% | 53,95% | 33,71% | 3,56% |
| Republicans | 2,37% | 1,30% | 55,65% | 19,17% | 1,86% |
| Mean | 1,52% | 2,15% | 54,80% | 26,44% | 2,71% |

# 6. Conclusion

Twitter is a digital environment where politicians in many countries around the world meet and communicate with their voters. At the same time, many people express their political views on Twitter. So, if we can analyze what these people wrote on Twitter and learn their political tendencies, can we predict the election results? More importantly, can we establish a relationship between the posts on Twitter and the political tendency? These are the questions that we and many researchers tried to answer before us. As a matter of fact, this issue has been researched since 2010, and studies conducted for many different countries show that there is a connection between political tendencies and political tendencies on Twitter, parties, politicians, elections, and election results can be predicted by using this link.

Political trend tracking based on Twitter data offers some advantages over polls. First of all, the audience whose opinions you take into account is much bigger than the polls. Estimates can be made by collecting the tweets of millions of people from Twitter. Therefore, the number of participants, ie the sample size, is much higher. In addition, it is much lower in cost and much faster. Collecting and analyzing millions of tweets from Twitter is a process that can be handled in a very short time after setting up the first model and with just a little bit of computer work.

The methods used for this work are various: That is, using some keywords such as the names and slogans of the parties and politicians, tweets about the elections or political tendencies are taken from Twitter. These tweets are categorized as positive, negative, and neutral by emotional analysis. There are different methods for this sentiment analysis. Machine learning can be used with artificial intelligence programs. There are also dictionary-based sentiment analysis tools. Afterward, a simple calculation can be made by proportioning the number of tweets about the parties categorized according to the emotion it contains. In other words, it is assumed that the higher the number of positive tweets about a party, the higher the chances of that party win.

In another method, with the assumption that people who tweet positively about the party support that party, calculations can be made by proportioning the numbers of people. However, although these methods can be used to predict the election results in previous studies, it is seen that they are not sufficient anymore. As a matter of fact, they do not consider it an important factor. That is the amount of tweet interaction. Those who are supporters of a party may tweet more than other party supporters and support their candidates. On the other hand, even if the other group does not tweet much, they can support the candidate by liking the tweets and re-tweeting (interaction).

For example, although the number of positive tweets about Clinton in the previous US election was higher than those about Trump, Trump won the election. Because the average number of likes for Trump tweets was much higher than Clinton [24]. This may indicate that Trump's supporters are less tweeting, meaning quieter than Clinton's supporters, but express their support for their candidates by liking the tweets that support Trump. Therefore, while following the political trend, not only the number of tweets, the number of people who posted these tweets, but also the interaction rates, likes, and retweets of these tweets should be taken into account.

According to what we have determined from the studies we had surveyed, the number of people tweeting for or against a party, the number of positive, negative, and neutral tweets about a party, the amount of retweeting of these tweets are important factors in the analysis of Twitter data for election results. In any of the studies in the literature, a prediction model that takes all these factors into account was not used. We developed a new selection outcome prediction model based on social media data that takes all these factors into account. We tested the validity of the analytical model we developed by trying to predict the outcome of the US November 2020 elections.

We estimated the USA 2020 November Election results using three alternative methods in the literature and the multi-factor method we developed. It is seen that the multi-factor method gives more accurate results than the other three alternative methods and even polls. The method we have developed offers an alternative approach to conventional tweet counting methods. With this new approach, it is thought that social media can be used to observe the change in perception about not only politicians but also companies, institutions, or individuals such as artists, sportsmen, and scientists.

# References

[1]   A. Bruns, S. Stieglitz, Towards more systematic Twitter analysis: metrics for tweeting activities, Int. J. Soc. Res. Methodol. 16 (2013) 91–108. https://doi.org/10.1080/13645579.2012.756095.

[2]   A. Tumasjan, T.O. Sprenger, P.G. Sandner, I.M. Welpe, Predicting elections with twitter: What 140 characters reveal about political sentiment, in: Fourth Int. AAAI Conf. Weblogs Soc. Media, Citeseer, 2010: pp. 455–479. https://doi.org/10.15581/009.37.2.455-479.

[3]   A. Livne, M.P. Simmons, E. Adar, L. a Adamic, The Party is Over Here : Structure and Content in the 2010 Election, October. 161 (2010) 201–208. https://doi.org/10.1007/s00024-003-2459-0.

[4]   T. Elghazaly, A. Mahmoud, H.A. Hefny, Political Sentiment Analysis Using Twitter Data, in: Proc. Int. Conf. Internet Things Cloud Comput., Association for Computing Machinery, New York, NY, USA, 2016. https://doi.org/10.1145/2896387.2896396.

[5]   P. Sharma, T.S. Moh, Prediction of Indian election using sentiment analysis on Hindi Twitter, Proc. - 2016 IEEE Int. Conf. Big Data, Big Data 2016. (2016) 1966–1971. https://doi.org/10.1109/BigData.2016.7840818.

[6]   P. Singh, R.S. Sawhney, K.S. Kahlon, Forecasting the 2016 US Presidential Elections Using Sentiment Analysis BT  - Digital Nations – Smart Cities, Innovation, and Sustainability, in: A.K. Kar, P.V. Ilavarasan, M.P. Gupta, Y.K. Dwivedi, M. Mäntymäki, M. Janssen, A. Simintiras, S. Al-Sharhan (Eds.), Springer International Publishing, Cham, 2017: pp. 412–423.

[7]   K. Jaidka, S. Ahmed, M. Skoric, M. Hilbert, Predicting elections from social media: a three-country, three-method comparative study, Asian J. Commun. 29 (2019) 252–273. https://doi.org/10.1080/01292986.2018.1453849.

[8]   B. Heredia, J.D. Prusa, T.M. Khoshgoftaar, Social media for polling and predicting United States election outcome, Soc. Netw. Anal. Min. 8 (2018) 1–16. https://doi.org/10.1007/s13278-018-0525-y.

[9]   W. Budiharto, M. Meiliana, Prediction and analysis of Indonesia Presidential election from Twitter using sentiment analysis, J. Big Data. 5 (2018) 51. https://doi.org/10.1186/s40537-018-0164-1.

[10]  Z. Xie, G. Liu, J. Wu, Y. Tan, Big data would not lie: prediction of the 2016 Taiwan election via online heterogeneous information, EPJ Data Sci. 7 (2018). https://doi.org/10.1140/epjds/s13688-018-0163-7.

[11]  K. Brito, N. Paula, M. Fernandes, S. Meira, Social Media and Presidential Campaigns – Preliminary Results of the 2018 Brazilian Presidential Election, in: Proc. 20th Annu. Int. Conf. Digit. Gov. Res., Association for Computing Machinery, New York, NY, USA, 2019: pp. 332–341. https://doi.org/10.1145/3325112.3325252.

[12]  R. Bose, R.K. Dey, S. Roy, D. Sarddar, Analyzing Political Sentiment Using Twitter Data BT  - Information and Communication Technology for Intelligent Systems, in: S.C. Satapathy, A. Joshi (Eds.), Springer Singapore, Singapore, 2019: pp. 427–436.

[13] M. Awais, S.U. Hassan, A. Ahmed, Leveraging big data for politics: predicting general election of Pakistan using a novel rigged model, J. Ambient Intell. Humaniz. Comput. (2019). https://doi.org/10.1007/s12652-019-01378-z.

[14] A. Bermingham, A.F. Smeaton, On Using Twitter to Monitor Political Sentiment and Predict Election Results, in: Proc. Ofthe Work. Sentim. Anal. Where AI Meets Psychol., Chiang Mai, Thailand, 2011: pp. 2–10.

[15] M. Bilal, S. Asif, S. Yousuf, U. Afzal, 2018 Pakistan General Election: Understanding the Predictive Power of Social Media, 12th Int. Conf. Math. Actuar. Sci. Comput. Sci. Stat. MACS 2018 - Proc. (2019) 1–6. https://doi.org/10.1109/MACS.2018.8628445.

[16] E.T.K. Sang, J. Bos, Predicting the 2011 dutch senate election results with twitter, in: Proc. Work. Semant. Anal. Soc. Media, 2012: pp. 53–60.

[17] M. Skoric, N. Poor, P. Achananuparp, E.-P. Lim, J. Jiang, Tweets and Votes: A Study of the 2011 Singapore General Election, in: 2012 45th Hawaii Int. Conf. Syst. Sci., IEEE, 2012: pp. 2583–2591. https://doi.org/10.1109/HICSS.2012.607.

[18] A. Ceron, L. Curini, S.M. Iacus, Using Sentiment Analysis to Monitor Electoral Campaigns: Method Matters—Evidence From the United States and Italy, Soc. Sci. Comput. Rev. 33 (2014) 3–20. https://doi.org/10.1177/0894439314521983.

[19] Q. You, L. Cao, Y. Cong, X. Zhang, J. Luo, A Multifaceted Approach to Social Multimedia-Based Prediction of Elections, IEEE Trans. Multimed. 17 (2015) 2271–2280. https://doi.org/10.1109/TMM.2015.2487863.

[20] A. Khatua, A. Khatua, K. Ghosh, N. Chaki, Can #Twitter-Trends predict election results? Evidence from 2014 Indian general election, Proc. Annu. Hawaii Int. Conf. Syst. Sci. 2015-March (2015) 1676–1685. https://doi.org/10.1109/HICSS.2015.202.

[21] A. Tsakalidis, S. Papadopoulos, A.I. Cristea, Y. Kompatsiaris, Predicting Elections for Multiple Countries Using Twitter and Polls, IEEE Intell. Syst. 30 (2015) 10–17. https://doi.org/10.1109/MIS.2015.17.

[22] R. Jose, V.S. Chooralil, Prediction of election result by enhanced sentiment analysis on twitter data using classifier ensemble Approach, Proc. 2016 Int. Conf. Data Min. Adv. Comput. SAPIENCE 2016. (2016) 64–67. https://doi.org/10.1109/SAPIENCE.2016.7684133.

[23] P. Chauhan, N. Sharma, G. Sikka, The emergence of social media data and sentiment analysis in election prediction, J. Ambient Intell. Humaniz. Comput. (2020) 1–27. https://doi.org/10.1007/s12652-020-02423-y.

[24] P. Grover, A.K. Kar, Y.K. Dwivedi, M. Janssen, Polarization and acculturation in US Election 2016 outcomes – Can twitter analytics predict changes in voting preferences, Technol. Forecast. Soc. Change. (2018) 1–23. https://doi.org/10.1016/j.techfore.2018.09.009.

[25] E. Abanoz, A Twitter-Based Analysis Of Hashtag And Mention Actions As An İndicator Of Turkish General Elections' Outcomes, Akdeniz Üniversitesi İletişim Fakültesi Derg. (2020) 73–90.

[26] Pew Research Center, Social Networking Popular Across Globe, 2012. https://www.pewresearch.org/global/2012/12/12/social-networking-popular-across-globe/.

[27] B. Bansal, S. Srivastava, On predicting elections with hybrid topic based sentiment analysis of tweets, Procedia Comput. Sci. 135 (2018) 346–353. https://doi.org/10.1016/j.procs.2018.08.183.

[28] M. Choy, M.L.F. Cheong, N.L. Ma, P.S. Koo, US Presidential Election 2012 Prediction using Census Corrected Twitter Model, Res. Collect. Sch. Inf. Syst. (2012) 1–12. http://arxiv.org/abs/1211.0938.

[29]   J. Ramteke, S. Shah, D. Godhia, A. Shaikh, Election result prediction using Twitter sentiment analysis, in: 2016 Int. Conf. Inven. Comput. Technol., IEEE, 2016: pp. 1–5.

[30]   A.J. Wicaksono, Suyoto, Pranowo, A proposed method for predicting US presidential election by analyzing sentiment in social media, in: Proceeding - 2016 2nd Int. Conf. Sci. Inf. Technol. ICSITech 2016 Inf. Sci. Green Soc. Environ., 2017. https://doi.org/10.1109/ICSITech.2016.7852647.

[31]   D. Anuta, J. Churchin, J. Luo, Election bias: Comparing polls and twitter in the 2016 us election, ArXiv Prepr. ArXiv1701.06232. (2017).

[32]   E. Sanders, A. van den Bosch, Optimising Twitter-based Political Election Prediction with Relevance andSentiment Filters, in: Proc. 12th Lang. Resour. Eval. Conf., 2020: pp. 6158–6165.

[33]   T. Mahmood, T. Iqbal, F. Amin, W. Lohanna, A. Mustafa, Mining Twitter big data to predict 2013 Pakistan election winner, in: Int. Multi Top. Conf., IEEE, 2013: pp. 49–54.

[34]   J.A. Cerón-Guzmánn, A Sentiment Analysis Model of Spanish Tweets. Case Study: Colombia 2014 Presidential Election, 2016.

[35]   P. Burnap, R. Gibson, L. Sloan, R. Southern, M. Williams, 140 characters to victory?: Using Twitter to predict the UK 2015 General Election, Elect. Stud. 41 (2016). https://doi.org/10.1016/j.electstud.2015.11.017.

[36]   R. Castro, L. Kuffó, C. Vaca, Back to #6D: Predicting Venezuelan states political election results through Twitter, in: 2017 4th Int. Conf. EDemocracy EGovernment, ICEDEG 2017, 2017. https://doi.org/10.1109/ICEDEG.2017.7962525.

[37]   L. Wang, J.Q. Gan, Prediction of the 2017 French election based on Twitter data analysis, in: 2017 9th Comput. Sci. Electron. Eng., IEEE, 2017: pp. 89–93.

[38]   M. Ibrahim, O. Abdillah, A.F. Wicaksono, M. Adriani, Buzzer Detection and Sentiment Analysis for Predicting Presidential Election Results in a Twitter Nation, Proc. - 15th IEEE Int. Conf. Data Min. Work. ICDMW 2015. (2016) 1348–1353. https://doi.org/10.1109/ICDMW.2015.113.

[39]   P. Singh, Y.K. Dwivedi, K.S. Kahlon, A. Pathania, R.S. Sawhney, Can twitter analytics predict election outcome? An insight from 2017 Punjab assembly elections, Gov. Inf. Q. (2020) 101444.

[40]   R. Jankowski, Predicting election polls results using machine learning tools, Pracownia Fizyki w Ekonomii i Naukach Społecznych, 2020. http://repo.bg.pw.edu.pl/index.php/en/r#/info/bachelor/WUT4a113c9f11ad427b98d0d9aa76c987be/.

[41]   D. Grimaldi, J. Diaz, H. Arboleda, Inferring the votes in a new political landscape. The case of the 2019 Spanish Presidential elections, Prepr. from Res. Sq. (2020). https://doi.org/10.21203/rs.3.rs-16463/v2.

[42]   A. Makazhanov, D. Rafiei, M. Waqar, Predicting political preference of Twitter users, Soc. Netw. Anal. Min. 4 (2014). https://doi.org/10.1007/s13278-014-0193-5.

[43]   H. Toker, S. Erdem, P. Özşarlak, 2015 Haziran ve Kasım Seçimlerinde Siyasal Eğilim: Yeni Bir Kamuoyu Ölçümleme Aracı Olarak Twitter, Erciyes İletişim Derg. 5 (2017) 96–116. https://doi.org/10.17680/erciyesakademia.291888.

[44]   Y. Wang, Y. Li, J. Luo, Deciphering the 2016 U.S. presidential campaign in the Twitter sphere: A comparison of the Trumpists and Clintonists, Proc. 10th Int. Conf. Web Soc. Media, ICWSM 2016. (2016) 723–726.

[45]   B. Yu, S. Kaufmann, D. Diermeier, Exploring the characteristics of opinion expressions for political opinion classification, (2008).

[46] G. Laboreiro, L. Sarmento, J. Teixeira, E. Oliveira, Tokenizing micro-blogging messages using a text classification approach, in: Proc. Fourth Work. Anal. Noisy Unstructured Text Data, 2010: pp. 81–88.

[47] F.Å. Nielsen, AFINN sentiment analysis in Python: Wordlist-based approach for sentiment analysis, Tech. Univ. Denmark. (2011). https://github.com/fnielsen/afinn.

[48] C.H.E. Gilbert, E. Hutto, Vader: A parsimonious rule-based model for sentiment analysis of social media text, in: Eighth Int. Conf. Weblogs Soc. Media (ICWSM-14). Available Http//Comp. Soc. Gatech. Edu/Papers/Icwsm14. Vader. Hutto. Pdf, 2014: p. 82.

[49] M. Thelwall, K. Buckley, G. Paltoglou, D. Cai, A. Kappas, Sentiment strength detection in short informal text, J. Am. Soc. Inf. Sci. Technol. 61 (2010) 2544–2558. https://doi.org/10.1002/asi.21416.

[50] A. Go, R. Bhayani, L. Huang, Twitter sentiment classification using distant supervision, CS224N Proj. Report, Stanford. 1 (2009) 2009.

[51] Allacronyms, Social Networking Acronyms and Abbreviations, (2020). https://www.allacronyms.com/social_networking/abbreviations (accessed September 3, 2020).

[52] B.I. Newman, J.N. Sheth, A model of primary voter behavior, J. Consum. Res. 12 (1985) 178–187.

[53] R. Redfield, R. Linton, M.J. Herskovits, Memorandum for the study of acculturation, Am. Anthropol. 38 (1936) 149–152.

[54] L. Shi, N. Agarwal, A. Agrawal, R. Garg, J. Spoelstra, Predicting US Primary Elections with Twitter, Opera Solut. (2012) 1–8.

[55] The Economist, Forecasting the US elections, (2020). https://projects.economist.com/us-2020-forecast/president (accessed December 1, 2020).

[56] M. Jockers, Package 'syuzhet'. URL: https://cran. r-project. org/web/packages/syuzhet (2007)

[57] L. Page, S. Brin, R. Motwani, T. Winograd, T. The PageRank citation ranking: Bringing order to the web. Stanford InfoLab (1999)

[58] T. Maehara, T. Akiba, Y. Iwata, K. I. Kawarabayashi, K.I. Computing personalized PageRank quickly by exploiting graph structures. Proceedings of the VLDB Endowment, 7(12) (2014), 1023-1034.

[59] A. B. Adcock, B. D: Sullivan, B. D., M. W. Mahoney, Tree decompositions and social graphs. Internet Mathematics, 12(5) (2016), 315-361.

[60] S. Yardi, D. Boyd, Dynamic debates: An analysis of group polarization over time on twitter. Bulletin of science, technology & society, 30(5) (2010), 316-327.

[61] F. Chen, Z. Chen, X. Wang, Z. Yuan, The average path length of scale free networks. Communications in Nonlinear Science and numerical simulation, 13(7) (2008), 1405-1410.

[62] A. Fronczak, P. Fronczak, J. A. Hołyst, Average path length in random networks. Physical Review E, 70(5) (2004), 056110.

[63] L. O. Prokhorenkova, E. Samosvat, Global clustering coefficient in scale-free networks. In International Workshop on Algorithms and Models for the Web-Graph (2014) (pp. 47-58). Springer, Cham.

[64] S. N. Soffer, A. Vazquez, Network clustering coefficient without degree-correlation biases. Physical Review E, 71(5) (2005), 057101.

[65] M. E. Newman, A measure of betweenness centrality based on random walks. Social networks, 27(1) (2005), 39-54.